\documentclass[prd,aps,amsmath,amssymb,letter,showpacs]{revtex4}
\usepackage{graphicx}
\usepackage{epsfig,rotate,latexsym}
\usepackage{hyperref}
\begin{document}
\title{Effects of quarks on the formation and evolution of Z(3)
walls and strings in relativistic heavy-ion collisions}
\author{Uma Shankar Gupta$^1$}
\email {guptausg@gmail.com}
\author{Ranjita K. Mohapatra$^2$}
\email {ranjita@iopb.res.in}
\author{Ajit M. Srivastava$^2$}
\email{ajit@iopb.res.in}
\author{Vivek K. Tiwari$^1$}
\email{vivek_krt@hotmail.com}
\affiliation{$^1$Physics Department, Allahabad University, 
Allahabad 211002, India \\
$^2$Institute of Physics, Sachivalaya Marg, Bhubaneswer 751005, India}

\begin{abstract}
 We investigate the effects of explicit breaking of Z(3) symmetry due 
to the presence of dynamical quarks on the formation and evolution of Z(3) 
walls and associated QGP strings within Polyakov loop model. We carry 
out numerical simulations of the first order quark-hadron phase
transition via bubble nucleation (which may be appropriate, for example, 
at finite baryon chemical potential) in the context of relativistic
heavy-ion collision experiments. Using appropriate shifting of the 
order parameter in the Polyakov loop effective potential, we calculate 
the bubble profiles using bounce technique, for the true vacuum as well 
as for the metastable Z(3) vacua, and estimate the associated
nucleation probabilities. These different bubbles are then nucleated 
and evolved and resulting formation and dynamics of Z(3) walls and 
QGP strings is studied. We discuss various implications of
the existence of these Z(3) interfaces and the QGP strings, especially
in view of the effects of the explicit breaking of the Z(3) symmetry
on the formation and dynamical evolution of these objects.
\end{abstract}

\maketitle

\section{INTRODUCTION}
\label{sec:intro}

The possibility of existence of topologically non-trivial structures
such as Z(3) interfaces and associated QGP strings in the quark-gluon
plasma phase \cite{gupta} is very exciting. In the context of 
relativistic heavy-ion collision experiments (RHICE), it provides the 
only system where domain walls and strings arise in a relativistic 
quantum field theory which can be investigated under laboratory control.
In earlier works \cite{gupta,z3str,znb} we have discussed various aspects of
existence of these objects in cosmology as well as in RHICE. These
topological objects arise in the high temperature deconfined phase of 
QCD due to spontaneous breaking of the Z(3) global symmetry of finite
temperature QCD, where Z(3) is the center  of the $SU(3)$ color 
gauge group  of QCD. Spontaneous breaking of Z(3) symmetry arises from
the non-zero expectation value of the Polyakov loop, $l(x)$, 
which is an order parameter for the confinement-deconfinement phase 
transition for pure gauge theory \cite{plkv}. The interpolation 
of $l(x)$ between three different degenerate $Z(3)$ vacua leads to the 
existence of domain walls (interfaces) together with topological strings 
when the three interfaces make a junction. We call these strings
as QGP strings \cite{gupta}. 

The properties and physical consequences of these $Z(3)$ interfaces have 
been discussed in the literature\cite{zn}. It has also been suggested  
that these interfaces should not be taken as physical objects in the 
Minkowski space \cite{smlg}.  Existence of these Z(3) vacua becomes
especially a non-trivial issue when considering the presence of
dynamical quarks. The effect of quarks on $Z(3)$ symmetry and $Z(3)$ 
interfaces etc. has been discussed in detail in the literature 
\cite{qurk1,qurk2}. It has been argued that the $Z(3)$ symmetry becomes 
meaningless in the presence of quarks \cite{qurk1}. 
Other view-point as advocated in 
many papers, asserts that one can take the effect of  quarks in terms 
of explicit breaking of $Z(3)$ symmetry \cite{qurk2,psrsk,psrsk2}, and we
will follow this approach. In this context we mention the recent work
of Deka et al. \cite{dglltc} which has provided a support for the existence of
these metastable vacua from Lattice. Although the temperatures
are high (close to 1 GeV) at which the indications of metastable vacuum are 
seen in ref.\cite{dglltc}, the important point is that these metastable 
Z(3) vacua seem to
exist at some temperature.  Since the presence of quarks lifts the 
degeneracy of different $Z(3)$ vacua \cite{qurk2,psrsk,psrsk2}, the 
$Z(3)$ interfaces become unstable and move away from the region with the 
unique true vacuum. Thus, with quark effects taken in terms of explicit
symmetry breaking, the interfaces survive as non-trivial topological
structures, though they do not remain solutions of time independent 
equations of motion. In our earlier investigations of these Z(3) walls 
and the QGP string we had neglected the effects of such an explicit 
symmetry breaking arising from quark effects \cite{gupta,z3str,znb}  and 
had investigated the properties and physical consequences of these 
objects $Z(3)$  in the context of  early universe  as well as in RHICE. 
In the present work, we will incorporate effects of explicit symmetry
breaking from quarks in the study of these objects.

Our numerical simulations in this work aim to investigate how the 
formation of $Z(3)$ walls and string network during the initial
confinement-deconfinement (C-D) transition in RHICE, and their subsequent 
evolution, gets affected by such explicit breaking of Z(3) symmetry. 
As in our earlier works, we model the pre-equilibrium stage of 
phase transition in our simulation as a quasi-equilibrium stage 
with an effective temperature which first rises (with rapid particle 
production) to a maximum temperature $T_0 > T_c$, where $T_c$ is the 
critical transition temperature, and then decreases due to 
continued expansion of plasma. 

In order to study the confinement-deconfinement (C-D) phase transition
in earlier works for the pure gauge case, we have been using the mean 
field effective potential of a polynomial form written in terms of the 
Polyakov loop expectation value $l(x)$ as proposed by Pisarski 
\cite{psrsk,psrsk2}. A linear term in $l(x)$ added to this effective 
potential in the mean field framework \cite{z3lnr,BankUK,GreeKarsh,Ogil} 
accounts for the explicit breaking of $Z(3)$ symmetry by the dynamical 
quarks whose presence act like a background magnetic field \cite{mgntc}.  
In our analysis in ref. \cite{gupta,znb} we had discussed the effects of
the explicit symmetry breaking term in view of the estimates of such
a term from ref. \cite{prsr}. We had 
found that the two degenerate vacua 
($l = e^{i2\pi/3}$,  and $l = e^{i4\pi/3}$), which get lifted with 
respect to the true vacuum ( with $l = 1$) on account of explicit 
breaking of $Z(3)$ symmetry in the QGP phase, have higher free energy 
than even the hadronic phase (with $l = 0$) at temperatures of order 200 
MeV. This does not seem reasonable because one would expect that any of 
the $Z(3)$ vacua which become meta-stable due to explicit symmetry breaking 
should still have lower free energy than the hadronic phase for values of 
temperature  $T >T_c$ enforcing that the system lies in the deconfining 
regime for such temperatures. In any case,
the estimates of \cite{prsr} refer to high temperature regime
and may not be applicable to temperatures near $T_c$. We thus use 
following considerations to constrain the magnitude of the strength
of the explicit symmetry breaking term. One approach can be to limit
it such that the metastable vacuum remains lower than the confining
vacuum for temperatures above $T_c$. We, however, limit explicit
symmetry breaking to further lower values by requiring that the first
order nature of the transition should remain at least in some 
range of temperatures above $T_c$.  

 We are using this first order transition model in the present work to 
discuss the dynamical details of quark-hadron transition, even though
the lattice results show that quark-hadron transition is most likely
a cross-over at zero chemical potential. The quark-hadron phase transition 
in the context of relativistic heavy-ion collision experiments is expected 
to be of first order for not too small values of the chemical potential 
which may be relevant for our study. Further, we are primarily interested 
in determining the time dependence of $Z(3)$ interfaces and string network 
structures, which result due to explicit breaking of $Z(3)$ symmetry during 
the phase transition. The formation of these objects is independent of the 
nature of phase transition as it results entirely due to finite 
correlation length in a fast evolving system, as shown by Kibble 
\cite{kbl}. The Kibble mechanism was first proposed for the formation of 
topological defects in the context of the early universe \cite{kbl}, but is 
now  utilized extensively for discussing topological defects production in 
a wide variety of systems from condensed matter physics to cosmology 
\cite{zrk}. Essential ingredient of the
Kibble mechanism is the existence of uncorrelated domains of the order
parameter which result after every phase transition occurring in finite
time due to finite correlation length. A first order transition allows
easy implementation of the resulting domain structure especially when
the transition proceeds via bubble nucleation. Keeping this view in mind,
we use the potential for Polyakov loop augmented with the addition of a 
linear term as in \cite{psrsk,psrsk2} to model the phase 
transition. Further we will be confining ourselves to the temperature/time 
ranges and such values of the coefficient of linear term in the effective 
potential that the first order quark-hadron transition proceeds via bubble 
nucleation.

The $Z(3)$ interfaces and strings will develop dynamics in the presence of 
explicit symmetry breaking and the interfaces will start moving
away from the direction where true vacuum exists. The strings will also not
have three interfaces forming symmetrically around it, and hence will 
start moving in some direction. Such motions may cause important 
differences on the long time behavior. Due to the quark effects, we will get
different nucleation probabilities/rates for the bubbles of meta-stable 
$Z(3)$ vacua and the true vacuum bubbles of the QGP phase. Meta-stable bubbles, 
being larger in size, may cover a larger fraction of the physical space and 
hence may lead to non-trivial consequences. The effects of quarks will be 
significant if a closed spherical wall (with true vacuum inside) starts 
expanding instead of collapsing. This effect may play an important role
in the early universe case because an expanding closed domain wall has to 
be large enough such that the surface energy contribution does not dominate 
over the volume energy. In the case of RHICE, the asymmetrical $Z(3)$ walls 
and associated strings will eventually melt away when the temperature drops 
below the deconfinement-confinement phase transition temperature $T_c$. 
However they will be leaving their signatures in the form of extended 
regions of energy density fluctuations 
(as well as $P_T$ enhancement of heavy-flavor hadrons \cite{apm} ). We will 
be estimating these energy density fluctuations which will lead to
multiplicity fluctuations. As in our earlier work \cite{gupta}, here also
our main focus will be in looking for the signals of extended regions of 
large energy densities in space-time reconstruction 
of hadron density. We mention here that a 
simulation of spinodal decomposition in Polyakov loop model has been 
carried out in ref. \cite{dumitru}, where fluctuations in the Polyakov loop
are investigated in detail. Our work, here and in ref.\cite{gupta} is 
focused on the formation of extended structures Z(3) walls, strings, and
extended regions of energy density etc., and in the present work, how
these are affected by effects of quarks.

       The paper is organized in the following manner. In section II, we 
briefly recall the Polyakov loop model of confinement-deconfinement phase 
transition and describe the effective potential proposed by Pisarski 
\cite{psrsk}. Here we discuss the effects of quarks in terms of
a linear term in the Polyakov loop in the effective potential, which 
leads to explicit breaking of the Z(3) symmetry. We discuss different 
estimates for the strength of this linear term in the context of
situations such that the transition is of first order. In section III, 
we discuss the effect of this term on the structure of Z(3) walls and 
strings, and on the structure of bubbles through which the phase transition
is completed. Here we describe our approach to extend the conventional 
technique of false vacuum decay to this case where different Z(3) bubbles have 
different profiles. What is of crucial importance to our discussion of the
formation of these objects is the nucleation rates of the bubbles of
different Z(3) vacua. Since these vacua are no more degenerate, the 
corresponding bubbles will in general have different nucleation rates.  
Section IV discusses nucleation rates for these different bubbles. One may 
expect that the metastable Z(3) vacua should be suppressed as the corresponding
bubbles have larger actions. We discuss the very interesting possibility
that despite having larger action the metastable vacua may have similar
(or even larger) nucleation rates as compared to the true vacuum.
This can happen when the pre-exponential factor dominates over the
exponential suppression term in the nucleation rate. This possibility 
is intriguing as the metastable vacua, being larger in size, may cover 
a larger fraction of the physical space and hence may dominate the 
dynamics of phase transition. 

Section V presents the 
numerical technique of simulating the phase transition via random 
nucleation of bubbles, which now have different sizes depending on the
corresponding Z(3) vacuum inside the bubble. Resulting domain walls
may show non-trivial behavior compared to the case without the quark
effects as a closed domain wall, enclosing the true vacuum, may expand
instead of contracting. Rough estimates, with our parameter choices,
show that this is expected when domain wall size exceeds about 50 fm.
The discussion of such a large physical region is more relevant in the 
context of the early universe and we plan to study this in a future work.
Here we will consider the case relevant to RHICE with lattice 
sizes of about (15 fm)$^2$  and study the effects of domain wall and 
string formation with temperature evolution as expected in a 
longitudinally expanding plasma. These results are presented in Sec.VI.
We also calculate the energy density fluctuations associated with 
$Z(3)$ wall network and strings, as in our earlier work \cite{gupta},
and discuss important differences for the present case with quark
effects. In section VII we discuss possible experimental signatures 
resulting from the presence of $Z(3)$ wall network and associate strings
especially including the effects of explicit symmetry breaking. 
Section VIII presents conclusions.

\section{THE POLYAKOV LOOP MODEL WITH QUARK EFFECTS}
\label{sec:plmdl}

We first briefly recall the Polyakov loop model for the confinement - 
deconfinement  phase transition. For the case of pure $SU(N)$ gauge 
theory, the expectation value of Polyakov loop $l(x)$ is the order 
parameter for confinement - deconfinement phase transition. 

\begin{equation}
  l(\vec{x}) ={1 \over N} Tr({\cal P} exp(ig\int^\beta_0 
A_0(\vec{x},\tau)d\tau))
\end{equation}

Where $A_0(\vec{x},\tau)$ is the time component of the vector potential 
$A_{\mu}(\vec{x},\tau)=A^{a}_{\mu}(\vec{x},\tau) T^a$, $T^a$ are the 
generators of $SU(N)$ in the fundamental representation,  $\cal{P}$ denotes 
path ordering in the Euclidean time $\tau$, g is the gauge coupling, and
$\beta = 1/T$ with T being the temperature. $N$ (= 3 for QCD) is the 
number of colors. The complex scalar field $l(\vec{x})$ 
transform under the global $Z(N)$ (center) symmetry transformation as 

\begin{equation}
 l(\vec{x}) \rightarrow {exp(2{\pi}in/N) l(\vec{x})}, \: n = 0,1,...(N-1)
\label{eq:zns}
\end{equation}

 The expectation value of $l(x)$ is related to $e^{-\beta F}$ where
$F$ is the free energy of an infinitely heavy test quark.
For temperatures below $T_c$, in the confined phase, the expectation value
of Polyakov loop  is zero corresponding to the infinite
free energy of an isolated test quark. (Hereafter, we will use the
same notation $l(x)$ to denote the expectation value of the Polyakov
loop.)  Hence the Z(N) symmetry is restored below $T_c$. Z(N) symmetry
is broken spontaneously above $T_c$ where $l(x)$ is non-zero
corresponding to the finite free energy of the test quark. Effective
theory of the Polyakov loop has been proposed by several authors
with various parameters fitted to reproduce lattice results for
pure QCD \cite{plkvall,psrsk,psrsk2}. We use the Polyakov loop effective 
theory proposed by Pisarski \cite{psrsk, psrsk2}. The effective 
Lagrangian density can be written as

\begin{equation}
L={N\over g^2} |{\partial_\mu l}|^2{T^2}- V(l)
\end{equation}

Where the effective potential $V(\it l)$ for the Polyakov loop, in case 
of pure gauge theory is given as

\begin{equation}
 V(l)=({-b_2\over2} |l|^2- {b_3\over 6}( l^3+( l^ \ast)^
 3)+\frac{1}{4}(|l|^2)^2){b_4{T^4}}
\end{equation}

At low temperature where $\it{l} = 0$, the potential has only one 
minimum. As temperature becomes higher than $T_c$ the Polyakov loop 
develops a non vanishing  vacuum expectation value $l_0$, and 
the $cos3\theta$ term, coming from the $l^3 + l^{*3}$ term above 
leads to $Z(3)$ generate vacua. Now in the deconfined phase, for a small
range of temperature above $T_c$, the $\it{l} = 0$ extremum becomes the local 
minimum (false vacuum) and a potential barrier exist between the local 
minimum and global minimum (true vacuum) of the potential.

To include the effects of dynamical quarks,  we will follow the approach 
where the explicit breaking of the $Z(3)$ symmetry is represented
in the effective potential by inclusion of a linear term in $l$
\cite{qurk2, psrsk, psrsk2, z3lnr}. The 
potential of Eq.(4) with the linear term becomes,

\begin{equation}
 V(l)=\Bigl( -\frac{b_1}{2}(l + l^*) -\frac{b_2}{2} |l|^2- 
{b_3\over 6} (l^3+ {l^{\ast}}^3) +\frac{1}{4}(|l|^2)^2 \Bigr){b_4{T^4}}
\label{eq:vb1l}
\end{equation}

Here coefficient $b_1$ measures the strength of explicit symmetry breaking. 
The coefficients $b_1$, $b_2$, $b_3$ and $b_4$ are dimensionless 
quantities. With $b_1 = 0$, other parameters $b_2$, $b_3$ and $b_4$ are
fitted in ref.\cite{psrsk,psrsk2,scav} such that 
that the effective potential reproduces the thermodynamics of pure 
$SU(3)$ gauge theory on lattice \cite{z3lnr,scav,latt}.  The coefficient 
$b_2$ is temperature dependent and given by

\begin{equation*}
 b_2(r) = \bigl(1 - \frac{1.11}{r}\bigr){\bigl(1 + \frac{0.265}{r}\bigr)}^2
{\bigl(1 + \frac{0.3}{r}\bigr)}^3 - 0.487 ;
\quad r = \frac{T}{T_c} ; \quad T_c =182~~ \text{MeV}
\end{equation*}

We use the value of temperature independent coefficients $b_3 = 2.0$ and 
$b_4 = 0.6061 \times \frac{47.5}{16}$. We choose the same value of $b_2$ for 
real QCD (with three massless quarks flavors). $b_4$ is rescaled by factor 
$\frac{47.5}{16}$ to incorporate extra degrees of freedom of QCD 
relative to pure $SU(3)$ gauge theory \cite{scav}. As temperature 
$T \rightarrow \infty$ the Polyakov loop expectation value approaches the 
value $x \sim b_3/2+\frac{1}{2} \sqrt{b_3^2+4b_2(T=\infty)}$ . To have the
normalization $\langle l(x) \rangle \rightarrow 1$ at $T \rightarrow \infty$,
the coefficients and field in the effective potential $V(l)$ in 
Eq.(\ref{eq:vb1l}) are rescaled as $b_1(T)\rightarrow b_1(T)/x^3$, 
$b_2(T)\rightarrow b_2(T)/x^2$, $b_3\rightarrow b_3/x$ and $b_4\rightarrow 
b_4x^4$, $l \rightarrow l/x$.

At temperatures above the critical temperature $T_c$ the potential $V(l)$ 
has three degenerate vacua in pure gauge theory (with $b_1 = 0$). The 
barrier heights between the local minimum ($l(x) = 0$) and the 
three global minima ($l=1, \rm{z}, \rm{{z}^2}$, corresponding to 
$\theta = 0, 2\pi/3, 4 \pi/3$) are all same. As the value of $b_1$ becomes 
non zero, the degeneracy of $Z(3)$ vacua gets lifted. Vacua corresponding 
to $\theta = 2 \pi/3$ ( $l= \rm{z}$) and $\theta = 4 \pi/3$ 
( $l= \rm{z}^2$) remain degenerate, with  energy which is higher than the 
$l= 1$ ($\theta = 0$) vacuum. Thus, $l = z$ and $l = z^2$ vacua become 
metastable and the $l = 1$ remains the only true vacuum (global minimum). 
Note that $l = z$ and $l = z^2$ are the two metastable vacua in the QGP
phase. Along with these, there is a metastable vacuum at $l = 0$
(for a small range of temperature above $T_c$)
which corresponds to the confining phase. 

  Estimates of explicit $Z(3)$ symmetry breaking arising from quark
effects have been discussed in the literature. In the high 
temperature limit, the  estimate of the difference in the potential
energies of the $l = z$ vacuum, and the $l = 1$ vacuum, $\Delta V$, is
given in ref. \cite{prsr} as,

\begin{equation}
 \Delta V \sim \frac{2}{3} \pi^{2} T^4 \frac{N_l}{N^3} (N^2 - 2)
\end{equation}

 where $N_l$ is the number of massless quarks. If we take $N_l = 2$ 
then $\Delta V \simeq 3 T^4$. At $T = 200$ MeV, the difference between 
the confining vacuum and the true vacuum from the effective potential in
Eq.(5) is about 150 MeV/fm$^3$ while $\Delta V$ from Eq.(6) at 
$T = 200$ MeV is about four times larger, equal to 
600 MeV/fm$^3$. As $T$ approaches $T_c$,
this difference will become larger as the metastable vacuum and the
stable vacuum become degenerate at $T_c$, while $\Delta V$ remains
non-zero.  It does not seem reasonable that at temperatures of order 
200 MeV (with $T_c = 182$ MeV for Eq.(5)) a QGP
phase (with quarks) has higher free energy than the hadronic phase. 
In any case, the estimates of Eq.(6)  were made in high temperature
limit and the extrapolation of these to $T$ near $T_c$ may be invalid.
We, thus, use different physical considerations to estimate the strength
of the explicit symmetry breaking term, i.e. the value of parameter
$b_1$ in Eq.(5), as follows.

 Note that as $b_1$ is increased from zero,  
the potential tilts such that the barrier between the metastable
confining phase and the true vacuum in the $\theta = 0$ direction 
decreases, resulting in the weakening of the first order phase transition. 
Finally, this barrier disappears for $b_1 \ge 0.11$ 
(at $T = T_c = 182$ MeV). For $b_1 \ge 0.11$ there is
no range of temperature where the phase transition is first order.
As we mentioned, our approach is to study the phase transition dynamics
via bubble nucleation. We thus choose a small value of $b_1 = 0.005$ such 
that the confinement - deconfinement phase transition is (weakly) first order 
phase transition for a reasonable range of temperature. 
The plot of the potential in $\theta = 0$ direction
for $b_1 = 0.005$ is shown in Fig.1 for $T = 200$ MeV. Note that
with $b_1 > 0$ the confining vacuum at $l = 0$ shifts towards positive
real value of $l$. With this value of $b_1$, the barrier between the confining
metastable vacuum and the true vacuum exists upto temperature $\simeq 225$ 
MeV which allows for a reasonable range of temperatures to discuss 
the bubble profiles and their nucleation probabilities. If we choose
larger values of $b_1$, the range of temperature allowing first
order transition becomes very narrow and formation and nucleation
of bubbles require fine tuning of time scale. 

\begin{figure*}[!hpt]
\begin{center}
\leavevmode
\epsfysize=5truecm \vbox{\epsfbox{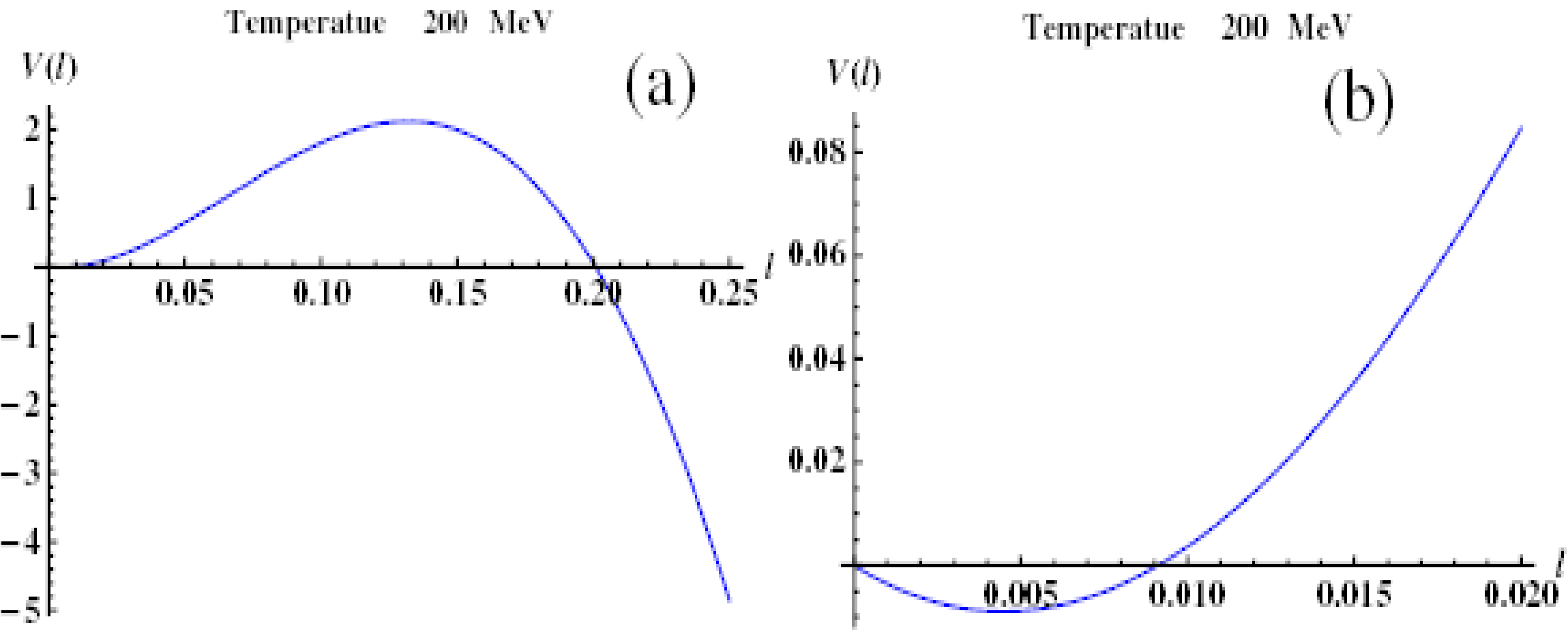}}
\end{center}
\caption{}{(a) Plot of  $V(l)$ (in MeV/fm$^3$) in $\theta = 0$ direction 
for $T = 200 MeV$ with $b_1 = 0.005$. (b) shows the plot near the origin,
showing that the confining vacuum has shifted
slightly from $l = 0$ towards $\theta = 0$ direction.}
\label{Fig.1}
\end{figure*}

 This, apparently ad hoc, procedure of fixing value of $b_1$ can be 
given a physical basis in the following way. Changing the value of 
coefficient $b_1$ changes the nature of phase transition from very strong 
first order to a very weak one. One can attempt to interpret it in the context
of QCD phase diagram, drawn in the plane of chemical potential ($\mu$) and 
temperature ($T$). The QCD phase transition is of strong first order for
large $\mu$, it becomes a weak first order transition with decreasing 
$\mu$, reaches to critical end point where the transition is of second 
order and then becomes crossover at lower $\mu$ values. If we 
assume that the effective potential in Eq.(5) (at least in form) can
describe these situations of varying chemical potential, then it looks
natural to assume that changing the value of $\mu$ is interpreted
in terms of changing the value of $b_1$ parameter in Eq.(5). Thus increasing
$\mu$ corresponds to lowering the value of $b_1$ making the phase
transition of stronger first order.  

 Note that the potential barrier between the confining vacuum and the 
true vacuum is maximum when $b_1$ is zero and the first order phase 
transition is strongest. This should correspond to the situation of 
largest $\mu$ according to the above argument, presumably corresponding to
the transition at very low temperatures in the QCD phase diagram.
However, with $b_1 = 0$ there is no explicit symmetry breaking. This will 
not be consistent with the expectation of explicit symmetry breaking 
arising from quark effects. Though one cannot exclude the possibility that 
the effects of dynamical quarks and that of net baryon number density 
may have opposite effects on the value of $b_1$, so that a strong first
order transition at large $\mu$ can be consistently interpreted in terms
of $b_1 = 0$. However, it is simpler to assume that even for the largest
value of $\mu$ (where the first order curve intersects the $\mu$ axis
in the QCD phase diagram), $b_1$ never becomes zero so that explicit
symmetry breaking remains present as expected. 

 Of course, it is clear that the parameter values used in Eq.(5), which 
were fitted using lattice results for $\mu = 0$ case, are no longer 
applicable, if non-zero values of $b_1$ are interpreted in terms of 
non-zero $\mu$. We will then need to assume that the required changes 
in the parameters of Eq.(5) for non-zero $\mu$ are not large. At the
very least we can say that, even if $b_1$ values we use here cannot
be justified, they help us capture some qualitative aspects of changes
in the formation and evolution of Z(3) walls and QGP strings when
quark effects are incorporated. 

\vspace{2.5cm}

\section{DOMAIN WALLS, STRINGS AND BUBBLES WITH EXPLICIT SYMMETRY BREAKING}
\label{sec:bbl}

 The explicit symmetry breaking arising from quark effects will have
important effects on the structure of topological objects; Z(3) walls
and the QGP strings. It will obviously also affect the nucleation
of bubbles of different Z(3) phases. First we qualitatively discuss
its effects on Z(3) walls and the QGP strings. For non-degenerate
vacua, even planar $Z(3)$ interfaces do not remain static, and move
away from the region with the unique true vacuum. Thus, while for the
degenerate vacua case every closed domain wall collapses, for the
non-degenerate case this is not true any more. A closed wall enclosing
the true vacuum may expand if it is large enough so that the surface
energy contribution does not dominate. Similarly it is no more possible
to have time independent solution for the QGP string. Without
explicit symmetry breaking a QGP string forming at the intersection of
three symmetrically placed Z(3) walls will be stationary. However,
with $b_1 \ne 0$ this is not possible for any configuration of domain
walls. In fact this type of situation has been discussed in the context
of early universe for certain types of anionic string models \cite{axion}.

 Apart from the structure of these objects, one also expects important
changes in the basic mechanism of formation of these objects during
phase transition. Without explicit symmetry breaking, these objects
will form via the Kibble mechanism, as discussed in detail in 
\cite{gupta}. In the presence of explicit symmetry breaking new
effects may arise as discussed in \cite{digal} where many string-antistring
pairs with small separations (which means small loops of strings 
or small closed domain walls in the present context) can form at the 
coalescence region of two bubbles.
This mode of production of topological objects arises from the
fluctuations of the order parameter and is entirely different from
the basic physics of the Kibble mechanism. As we are using very small
value of explicit symmetry breaking, we do not expect this
new mechanism to play an important role here. However, for larger
values of $b_1$, this production mechanism may play an important role
in determining the Z(3) wall and string network resulting from a
first order QCD phase transition.

 General picture of the formation of these objects during first order
QCD transition via bubble nucleation was described in detail in ref.
\cite{gupta} for the case without explicit symmetry breaking and we 
briefly summarize it below. Subsequently we will discuss the effects
of explicit symmetry breaking on the bubble profiles, their nucleation
rates, and on general dynamics of the phase transition.

We calculate the bubble profile of QGP phase using Coleman's 
technique of bounce solution \cite{clmn} for true vacuum ($l =1$) and 
for metastable vacua ($l = z, z^2$). We seed these bubbles in the false 
(hadronic) background randomly with their nucleation rates calculated 
at an appropriate value of temperature  $T > T_c$ (such
that the nucleation rate is appreciable). The value of the phase of the
complex order parameter $l$ is constant inside a given bubble (to minimize
the free energy), while it changes from one bubble to another 
randomly (corresponding to the choices of three vacua). The variation
of the orientation of the order parameter from one bubble to another
provides the essential ingredient of the Kibble mechanism leading to
a domain structure and formation of topological objects at the
intersection of domains. We evolve this initial field configuration 
with the equations of motion using leap frog algorithm.  Bubbles grow with 
time and coalesce with each other.  The bubbles with same vacuum merge 
together to form a bigger region of same vacuum while the bubbles with 
different vacua remain separated by a wall/interface of high energy density
after coalescence. These are the Z(3) domain walls. These domain walls 
are solutions of field equations of motion, interpolating between different 
Z(3) vacua, and survive till very long time as QGP evolves. Eventually,
either walls collapse/merge away, or they melt as the temperature of
expanding QGP falls below $T_c$ and Z(3) symmetry is restored.

 Spontaneous breaking of Z(3) symmetry in the QGP phase leads to
three different topological domain walls separating the three different 
$Z(3)$ vacua. The intersection point of the three domains walls 
leads to a topological string (the {\it QGP} string) which was discussed 
in detail in ref. \cite{z3str}. This string arises as the order
parameter $l$ completes a closed loop around $l=0$ in the complex $l$ space
when one encircles the intersection point of the three domain walls in 
the physical space \cite{z3str}. Thus, these are topological strings which
exist in the QGP phase and have confining core (with $l = 0$).
As bubbles of different Z(3) vacua coalesce with each other, a network
of Z(3) walls forms and at the intersection of Z(3) walls, QGP strings form.
A detailed investigation of this for the case without explicit symmetry 
breaking (i.e. $b_1 = 0$), using 2+1 dimensional simulation representing 
the central rapidity region, was carried  out in ref. \cite{gupta}.

 The above picture of the dynamics of bubble nucleation, coalescence, and
formation and evolution of Z(3) walls and QGP strings will be affected by
the presence of explicit symmetry breaking in important ways. With $b_1 \ne 0$,
the three $Z(3)$ vacua are no longer degenerate. The two vacua ($l=z,
z^2$) corresponding to $\theta = 2\pi/3, 4\pi/3$ get lifted and become 
metastable. Only the third one with real expectation value of $l$ 
remains stable. The energy difference between the confining vacuum (near
$l = 0$, note that due to $b_1 \ne 0$, the confining vacuum shifts slightly) 
and the two metastable Z(3) vacua (with $l = z,z^2$) is smaller than the 
energy difference between the confining vacuum and the true vacuum.
This leads to larger bubble size for metastable vacuum than the bubble 
of true vacuum, with larger value of associated action (free energy). 
The energy difference between the confining vacuum and true or metastable 
vacuum increases with increase in temperature so the bubble sizes 
decreases with increase in temperature.

In the non-degenerate case  with non vanishing explicit symmetry breaking 
the false vacuum of potential gets shifted towards real axis by an small 
amount $\epsilon$. This shift is minimum for temperature closer to $T_c$ 
and increases as we increase the temperature. Further, the local maximum 
of the potential barrier and the metastable vacua are not in same direction 
but there is a small angular shift between them. These aspects make it
difficult to apply the Coleman's technique of bounce solution of a scalar
field for the present case as we will discuss below. First we review the 
basic features of the first order transition via bubble nucleation.

 A first order phase transition proceeds by the nucleation of a true 
vacuum bubble in the background of false vacuum. A true vacuum bubble 
produced, will grow or collapse depending on the free energy change of the 
system. The change in the free energy of the system because of the 
creation of a true vacuum bubble of radius R is 

\begin{equation}
F(R) = F_s + F_v = 4\pi R^2 \sigma- {4\pi \over 3} R^3 \eta
\end{equation}

Here $F_v$ is the volume energy and $F_s$ is the surface energy of the bubble. 
For a strong first order phase transition, one can analytically determine 
the potential energy difference $\eta$ between the confining vacuum and 
relevant Z(3) vacuum and the surface tension $\sigma$ from bounce 
solution (at least for a scalar field). Minimization of this free energy 
determines the critical radius $R_c = \frac{2\sigma}{\eta}$. The volume 
energy of the bubbles with  radius $R > R_c$ dominates over its surface 
energy and the bubbles expand to transform the false vacuum to true vacuum. 
The smaller bubbles ($R < R_c$) for which surface energy dominates over 
the volume energy, shrink and disappear. For strong first order transition,
calculation of $\eta$ and $\sigma$ separately can be done as one is
dealing with the thin wall bubbles where the bubble size is much larger
than the thickness of the bubble wall, so that there is clear separation 
between the bubble core and the bubble wall. For the parameter values, and
the temperature range of our interest, we will be dealing with thick wall
bubbles where bubble size if of the same order as the bubble wall. For this
purpose, the expression in Eq.(7) is not of use, and one has to calculate 
the bubble profile numerically using Coleman's technique of bounce solution 
and determine its action to calculate nucleation probabilities. 

The theory of semiclassical decay of false vacuum at zero temperature is 
given in ref \cite{clmn} and its extension to finite temperature was given 
in ref \cite{linde}. The Coleman's technique is applicable for real scalar 
field. To calculate bubble profile for complex scalar field $l$ (with
$b_1 = 0$), in ref. \cite{gupta}, the phase angle $\theta$ was taken 
to be constant  by fixing it in the direction of the relevant Z(3)
vacuum, i.e. $\theta = 0, 2\pi/3, {\rm or,} 4\pi/3$. This reduced the
problem again to a real scalar field calculation and Coleman's technique
could be directly applied. (However, there are important issues for the case
of complex scalar field regarding the calculation of nucleation rates 
which require calculation of determinant of fluctuations around the bounce 
solution. A brief discussion of these issues is provided in ref. \cite{gupta}.

We calculate the bubble profile in $3+1$ dimension. However, we evolve it only
by the $2+1$ dimensional field equations. This is because of rapid 
longitudinal expression which simply stretches the bubbles in the
longitudinal direction, while its transverse evolution proceeds according
to field equations.  We neglect the transverse expansion of system which
is certainly a good approximation during early stages of bubble nucleation
(during initial transition from confining phase to the QGP phase with
time scales of order 1 fm). At finite temperature, the 3+1 - dimensional 
theory will reduce to an effectively 3 Euclidean dimensional theory if 
the temperature is sufficiently high, which we will take to be the 
case \cite{gupta}.  For this 3 dimensional Euclidean theory, the bubble 
profile is the solution of the following equation

\begin{equation}
 \frac{d^{2}l}{dr^{2}} + \frac{2}{r} \frac{dl}{dr} =
\frac{g^2}{2NT^2} \frac{\partial{V}}{\partial{l}}
\label{eq:eom}
\end{equation}

where $r =r_E = \sqrt{{\vec{x}}^2 +{t_E}^2}$, subscript E denotes 
coordinates in the 3 dimensional Euclidean space. We use fourth order 
Runge-Kutta method to solve Eq.(\ref{eq:eom}).  
For $b_1 = 0$ , the relevant boundary 
conditions on $l$ to calculate the bubble profile are $l = 0$ as 
$r \rightarrow \infty$ and $\frac{dl}{dr} = 0$ at $r = 0$. However,
with $b_1 > 0$ this is no longer applicable. This is because with
$b_1 \ne 0$ the confining vacuum is shifted from $l = 0$ along 
$\theta= 0$ direction by an amount $\epsilon$. We calculate the bubble 
profile at $T = 200$ MeV and at this temperature $\epsilon = 0.0045$
(see, Fig.1b). We thus re-write the effective potential in Eq.(5) in terms 
of a shifted field $l^\prime = l - \epsilon$. In  terms of $l^\prime$ the 
confining vacuum again occurs at $l^\prime = 0$ and the standard boundary
conditions as discussed above can be applied for solving Eq.(8) for
the bounce solution. Hereafter all discussion will be in terms of
this shifted field $l^\prime$ which, for simplicity we will denote 
as $l$ only. 
   
  Another complication occurs in calculating the bubble profile for
the metastable Z(3) vacua. The earlier technique for $b_1 = 0$ case
of simply fixing $\theta = 2\pi/3$ or $\theta = 4\pi/3$ for the
two respective vacua, thereby reducing the problem to a real scalar 
field case, cannot be applied here directly. This is because with
$b_1 \ne 0$, the maximum of the respective potential barrier and 
direction of the corresponding metastable vacuum are not in the same 
direction (due to the tilt of the potential resulting from $b_1 \ne 0$).
However, the difference between the two directions, i.e. between the
$l = z$ vacuum and the direction of the top of the corresponding
barrier, is very small, of order $\theta = 0.9^\circ$. Same is true for
$l = z^2$ vacuum. We then fix $\theta$ along $l=z$ and $l = z^2$ vacua
respectively to get the approximately valid bubble profile using Eq.(8). 
(Both these directions differ slightly from $\theta = 2\pi/3$
and $\theta = 4\pi/3$ now. Note again all this is using the shifted
field which we are again denoting as $l$.) Recall, that we are 
calculating 3+1 dimensional critical bubble and evolving it by 2+1 
dimensional equations with the bubble becoming supercritical for 2+1 dim. 
equations \cite{gupta}.  Further we are studying the situation of 
rapidly changing temperature. Thus exact profile of the critical 
bubble at the nucleation time if not of much relevance.   

 As we had mentioned above, we choose the value of $b_1$ such that the 
barrier between the confining vacuum and various Z(3) vacua remains 
non-zero up to some range of temperature so that bubble formation
can be carried out. We choose $b_1 = 0.005$ with which the
barrier between the confining vacuum and true vacuum exist upto 
temperature $\simeq 225$ MeV. The first order phase transition via
bubble nucleation is possible only upto this temperature.

\begin{figure*}[!hpt]
\begin{center}
\leavevmode
\epsfysize=7truecm \vbox{\epsfbox{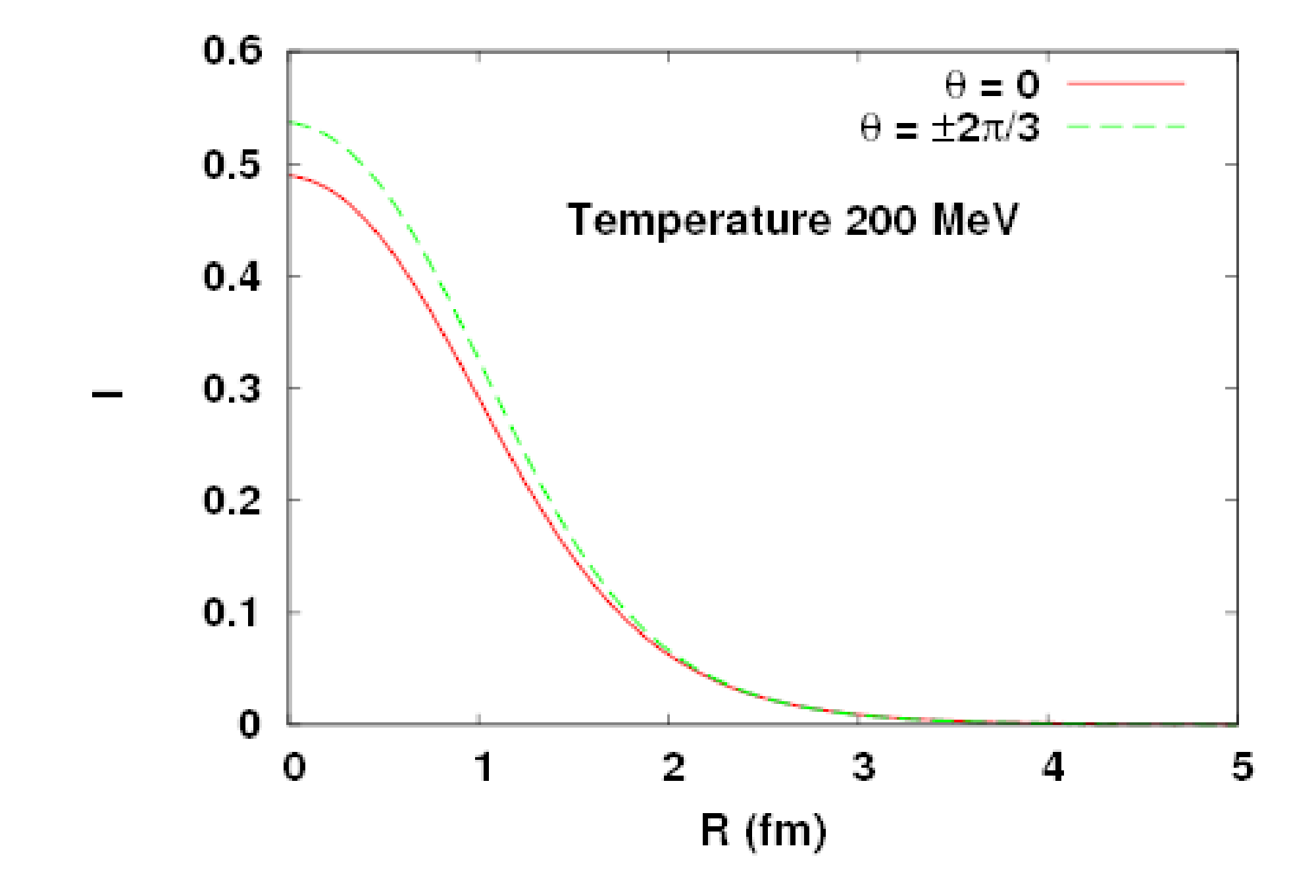}}
\end{center}
\caption{}{Critical bubble profiles for the different Z(3) vacua
for $b_1 = 0.005$.}
\label{Fig.2}
\end{figure*}

\section{NUCLEATION RATES FOR DIFFERENT BUBBLES}
\label{sec:nuclr}

For the finite temperature case, the tunneling probability per unit
volume per unit time in the high temperature approximation is
given by \cite{linde} (in natural units)

\begin{equation}
 \Gamma = A~e^{-S_3(l)/T} 
\end{equation}

where $S_3(l)$ is the 3-dimensional Euclidean action for the  Polyakov
loop field configuration that satisfies the classical Euclidean equations
of motion. The dominant contribution to the exponential term in $\Gamma$ 
comes from the bounce solution which is the least action $O(3)$ symmetric 
solution of Eq.(8). For a theory
with one real scalar field in three Euclidean dimensions the
pre-exponential factor arising in the nucleation rate of critical
bubbles has been estimated, see ref. \cite{linde}. The pre-exponential
factor obtained from \cite{linde} for our case becomes

\begin{equation}
 A =  T^4 \left({S_3(l) \over 2\pi T}\right)^{3/2}
\end{equation}

 As emphasized in \cite{gupta}, the results of \cite {linde} were
for a single real scalar field and one of the crucial ingredients
used in \cite {linde} for calculating the pre-exponential factor was
the fact that for a bounce solution the only light modes contributing to
the determinant of fluctuations were the deformations
of the bubble perimeter. Even though we are discussing the case of
a complex scalar field $l(x)$, this assumption may still hold as we
are calculating the tunneling from the false vacuum to one of the
Z(3) vacua.  This assumption may need to be revised when
light modes e.g. Goldstone bosons are present which then also have
to be accounted for in the calculation of the determinant.

 A somewhat different approach for the pre-exponential factor in Eq.(9)
is obtained from the nucleation rate of  bubbles per unit volume for a 
liquid-gas phase transition as given in ref. \cite{langer,csern}.
In ref.\cite{gupta} we had considered these estimates for the nucleation
rate as well as those obtained from Eq.(10). It was found that for the 
parameter values in Eq.(5) and for the temperature/time scales relevant
for RHICE, the nucleation rates obtained using the liquid-gas transition
approach of ref.\cite{langer,csern} were completely negligible such
that even nucleation of one bubble of QGP phase was not likely at RHICE. 
As one needs several bubbles to discuss the formation of
Z(3) walls and strings, these estimates clearly cannot be used here.
As in ref.\cite{gupta} we will follow the approach based on Eq.(10)
for our case which gave reasonable nucleation rates leading to
the possibility of formation of several bubbles for the case
of RHICE. We may mention here that for nucleation of bubbles of
the Polyakov loop $l$ it may anyway be better to use a field theory
approach as in ref.\cite{linde} rather than the approach of ref.
\cite{langer,csern} which is more suitable for the description
of phase transition in terms of plasma degrees of freedom. Though,
the parameters of Eq.(5) have been fitted with lattice QCD, it
is still not very clear whether the bubbles should be viewed in terms
of an order parameter field representing some background condensate
(as the Polyakov loop $l$), or just different phases of an interacting
plasma.

  We thus proceed with the calculation of  nucleation rates of the bubbles
using Eqs.(9),(10). Fig.2 shows the profiles of the bubbles for $l = 1$ and
$l = z$ vacua at $T = $ 200 MeV ($l = z^2$ bubbles has the same profile 
as the $l = z$ bubble). We note that $l = z$ bubble is somewhat larger
as expected. Using such bubble profiles we calculate the respective values of
the action $S_3$ and estimate the nucleation rates for metastable and true 
vacuum at different temperatures. To calculate number of bubbles for a 
typical nucleus nucleus collision, we consider a circle of 8 fm radius in 
transverse plane with 1 fm thickness in the longitudinal direction. The bubble 
nucleation for 1 fm time obtained from the nucleation rate given
in Eqs.(9),(10) leads to about 3 - 5 bubbles in this region. (The approach 
followed in ref. \cite{csern} gives the nucleation rate of about
$10^{-4}$ fm$^{-4}$ in the relevant temperature range, leading
to negligible nucleation of bubbles). 

 One may expect that nucleation rate of the two metastable Z(3) vacua will
be smaller than that of the true Z(3) vacuum due to larger action
$S_3$ of the metastable vacuum leading to exponential suppression. 
However, here we see an interesting interplay between the exponential 
factor $e^{-S_3(T)/T}$ (Eq.(9)) and the prefactor $A$ as given in
Eq.(10). If $S_3(T)$ is much larger than $T$ then the nucleation rate is
dominated by the exponential factor confirming the above expectation. Thus, 
the nucleation rate of metastable vacuum bubble is much smaller than
the true vacuum bubble when temperature is closer to $T_c$. The nucleation 
rate of true vacuum bubble and metastable vacuum bubble at temperature near 
$T_c$ (at $T = $ 185 MeV) is of the order of $\sim 1.3 \times 10^{-5}$ 
fm$^{-4}$ and $\sim 3.4 \times 10^{-7}$ fm$^{-4}$ respectively.  As we 
increase the temperature from $T_c =182$ MeV, the nucleation rate of 
metastable vacuum bubble increases and becomes almost equal to that of 
true vacuum bubble at $T \simeq 200$ MeV (both rates being about
$\sim 2.4 \times 10^{-2}$ fm$^{-4}$).
This happens because at these temperatures $S_3 \simeq T$ so that the 
decrease of the exponential term for a larger $S_3$ (corresponding to
the metastable vacuum) is not very significant. However, the pre-exponential
factor $A$ in Eq.(10) increases with $S_3$ and this increase of the 
prefactor term starts dominating the exponential factor in the nucleation 
rate equation for $T \ge 200$ MeV.  
For larger temperatures, the nucleation rate for 
metastable vacuum bubbles become larger than the true vacuum  bubbles. The 
nucleation rates of metastable and true vacuum bubble at temperature $215$ 
MeV are the order of $\sim 1.5 \times 10^{-2}$ fm$^-{4}$ and $7.7 \times 
10^{-3}$ fm$^{-4}$ respectively. At higher temperatures though the nucleation 
rate for both bubbles decrease but the metastable bubble nucleation rate 
remains larger. This result is very interesting as it shows that at suitable
temperatures the metastable Z(3) vacua will have larger nucleation rate than
the true Z(3) vacuum. Further, these metastable vacuum bubbles are also of 
larger size than the bubble of true vacuum. Thus one may expect a larger
fraction of the QGP region to end up in the metastable Z(3) vacuum regions
after the phase transition which may have interesting implications.
For example, we will see below that the metastable vacuum bubble walls 
have much higher concentration of energy density than the true vacuum bubble
walls.  We will use $T = 200$ MeV for the bubble nucleation as the nucleation 
rate are same for both the true vacuum and metastable vacuum bubbles.

\section{NUMERICAL TECHNIQUES}
\label{sec:ntech}

In this work, we carry out a $2+1$ dimensional field theoretic simulation 
of the formation and evolution of QGP phase bubbles representing the
central rapidity region of QGP in RHICE. Bubbles are nucleated randomly in
the confining background.  We calculate the bubble profiles in $3+1$ dimension
and use these profiles for the evolution in 2+1 dimensions. As explained 
above, this represents transverse evolution of these bubbles by field
equations and their longitudinal evolution is simply given by the Bjorken
longitudinal expansion \cite{bjorken}.  We nucleate bubbles at the 
temperature $200$ MeV at which the metastable and true vacuum 
bubbles have the nucleation rates of the same order $\simeq 0.024$ fm$^{-4}$ 
so that the number of metastable and true vacuum bubbles seeded remains 
almost equal. Initially the the field $l(\vec{x})$ is zero every where and 
bubbles of QGP phase are  nucleated over the whole lattice with random choice 
of their location. (Again, recall that we are using the shifted field here 
with $b_1 \ne 0$).  Bubbles are nucleated with the condition that one bubble 
should not overlap with the other. We implement this condition by checking 
that whether the region where bubble is going to be nucleated, lies in the 
false vacuum or not. If in the region a bubble has seeded already, the 
next bubbles will be seeded at some other random position with same conditions. 
(These techniques for the formation and evolution of bubbles in a first order
transition are the same as used in ref.\cite{ajit}.) 

 We take the initial temperature of the system to be zero (representing initial
confining system) and  it is taken to increases linearly with time up to 
$T = $400 MeV, at (proper) time $\tau = \tau_0 = 1$ fm.  The bubble nucleation 
is possible only in the range of temperature where the transition is of first 
order. The barrier in between false vacuum and true vacuum as well as false 
vacuum and metastable vacua of Eq.(\ref{eq:vb1l}) exist only for the 
temperature $T = 182$ MeV to $T \simeq 225$ MeV for our chosen value of 
$b_1 = 0.005$. The nucleation of bubbles is possible 
only during the time when temperature 
linearly increases from $T = T_c = 182$ MeV to $T \simeq 225$ MeV. In order to 
have a reasonable range of temperatures for bubble nucleation and evolution 
we nucleate bubbles at $T = $ 200 MeV. Note that bubbles should also
be nucleated at higher temperatures, say near $T = $ 225 MeV. These will be
smaller in size. Along with such bubbles there will also be subcritical bubble 
which shrink  fast and disappear due to the surface energy domination. Such 
bubbles should be incorporated to account for fluctuations \cite{ajit}, but 
we will ignore these here.

In Relativistic Heavy Ion Collision Experiments the QGP bubbles are 
nucleated in the hadronic phase during the time span when temperature changes 
from the transition temperature to the maximum temperature $T_0 = 400$ MeV in
the pre-equilibrium stage, hence this should lead 
to the presence of metastable and true 
vacuum bubbles of different sizes at a given time. These bubbles expand in 
hadronic background with time and ultimately the whole system gets 
converted to the QGP phase. We choose to seed the bubbles at a fix nucleation 
temperature because the QGP bubbles being nucleated in hadronic background 
have zero velocity initially and remain almost static during the remaining 
pre-equilibrium time $ \simeq 0.5$ fm when temperature increases from $T = T_c 
= 182$ MeV to $T = T_0 = 400$ MeV. The growth of bubbles nucleated at different 
time and the increase in their velocity until the temperature reaches to 
$400$ from the nucleation temperature, are negligible in this short time 
span. Therefore our choice for simplicity to seed bubbles at fixed temperature 
is a reasonable approximation. We choose $T = 200 $ MeV as at this  
temperature the true vacuum and the metastable vacuum bubbles have almost equal 
nucleation rate and both kind of bubbles are possible with equal probability. 
This provides us a better opportunity to study the dynamics of metastable 
vacuum bubbles together with that of true vacuum bubbles and its
effect on true vacuum bubbles evolution. 

After nucleation, bubbles are evolved 
by time dependent equation of motion in the Minkowski space \cite{rndrp}

\begin{equation}
  \frac{\partial^{2}l_j}{\partial\tau^{2}} + \frac{1}{\tau} 
\frac{\partial l_j}{\partial \tau}  -\frac{\partial ^{2}l_j}{\partial x^{2}}   
-\frac{\partial^{2}l_j}{\partial y^{2}} 
= -\frac{g^2}{2NT^2} \frac{\partial{V(l)}}{\partial{l_j}} ;
\quad j = 1, 2
\label{eq:evolution}
\end{equation}

with $\frac{\partial l_j}{\partial \tau} = 0$ at $\tau = 0$ and $l = 
l_1 +i l_2$.

We take a $2000 \times 2000$ lattice with physical size 16 fm x 16 fm. (as 
appropriate for, say Au-Au collision at RHICE). We take this lattice as the 
transverse plane of the QGP formed in a central collision and consider 
the longitudinal extension of 1 fm in the mid rapidity region. The evolution 
of metastable and true vacuum bubbles with different $Z(3)$ vacuum inside 
gives rise to the domain wall and string networks. The domain walls form when 
the two bubbles of different $Z(3)$ vacua coalesce with each other. The 
intersection of three domain walls forms a string. In our simulation these
objects are formed in the transverse plane. Hence, the domain walls appear 
as curves while the cross section of three dimensional strings appear as  
vortices.

In the relativistic heavy ion collision, the thermalization time for a 
Au - Au collision at $200$ MeV is expected to be $\tau \le 1$ fm time. As 
mentioned above, we model the system in our simulation such that there is a 
linear increase in temperature in the pre-equilibrium stage. It starts from 
$T = 0$ and reaches to a maximum value of $T = 400$ within time $\tau = 0$ 
to $\tau = \tau_0 = 1$. After that it decreases according to Bjorken's 
scaling due to the continued expansion in longitudinal direction
\cite{bjorken}

\begin{equation}
T(\tau) = T(\tau_0) \left({\tau_0 \over \tau}\right)^{1/3}
\label{eq:bjscal}
\end{equation}

In our numerical simulation we evolve the field using the periodic, fixed, 
and free boundary conditions for the square lattice. We  present our results 
for the free boundary condition case where the field (waves) crossing the 
boundary during evolution go out permanently. This condition minimizes 
effects due to boundary points in the evolution of field (field reflection 
from boundary points in fixed boundary condition and mirror reflection as in 
periodic boundary condition). We use additional dissipation in a thin
strip of ten points near the boundary to reduce the (minor) boundary effects 
in the use of free boundary conditions. For representing the situation of 
heavy ion collision experiments we nucleate bubbles within a circular region 
of 8 fm radius on the lattice of physical size 16 fm x 16 fm. With
$\Delta x = 0.008$ fm, we use $\Delta t = \Delta x/\sqrt 2$  and $\Delta t = 
0.9 \Delta x/\sqrt 2$ to satisfy the Courant stability criteria. The stability 
and accuracy of the simulation is checked using the conservation of 
energy during simulation. The total energy fluctuations remains few percent 
without any net increase or decrease of total energy in the absence of 
dissipative $\dot l$ term in Eq.(\ref{eq:evolution}) as well as any other 
dissipation for periodic and fixed boundary condition. 

\section{RESULTS OF THE SIMULATION}
\label{sec:results}

 General picture of the phase transition remains similar to the case of 
$b_1 = 0$ discussed in \cite{gupta}, but there are important differences. 
We show in Fig. 3 the various stages of
the formation and evolution of different Z(3) bubbles and subsequent formation
and evolution of Z(3) walls and strings. In order that one can compare
with the case discussed in ref.\cite{gupta} for $b_1 = 0$ case, we
present in Fig.3 the case of 5 bubbles in a  16 fm $\times$ 16 fm region,
similar to the case discussed in ref.\cite{gupta}.
Fig.3a shows the initial plot of $l(x)$ showing nucleation of 5 bubbles
at $\tau = 0.5 $ fm.
Fig.3b shows the plot of $l(x)$ at $\tau = 1.5$ fm showing the expansion of 
bubbles. Fig.3c shows the plot of the phase of $l(x)$ at the initial stage 
and Fig.3d shows the phase plot at $\tau = 3.2$ fm showing clearly the
formation of domain walls and a QGP string near (x=8 fm,y=9 fm).
The important difference in the dynamics of true vacuum bubbles and the 
metastable vacuum bubbles can be seen in the surface plots of energy density 
(in GeV/fm$^3$) at $\tau = 0.75 $ fm (Fig.3e) and at $\tau = 2.6$ fm (Fig.3f).
Note that in all the figures we plot energy density in GeV/fm$^3$ as we
are considering the central rapidity region with thickness of about 1 fm.
With similar energy densities to begin with, by the time $\tau = 2.6$ fm, the 
energy density at the walls of bubbles of true vacuum is much smaller
than the energy density of walls for the false vacuum bubbles.

\begin{figure*}[!hpt]
\begin{center}
\leavevmode
\epsfysize=7truecm \vbox{\epsfbox{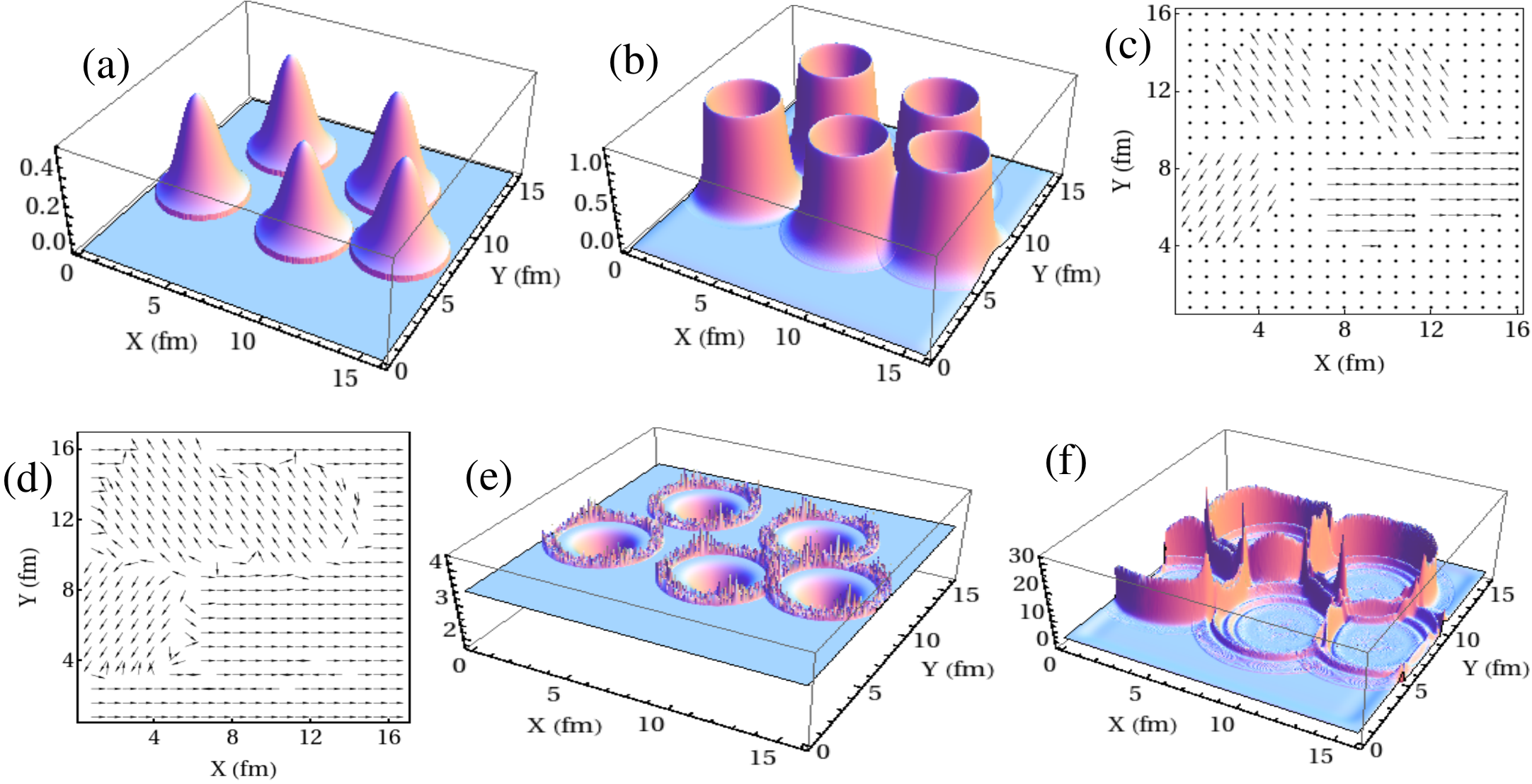}}
\end{center}
\caption{}{(a) and (b) show plots of profiles of $l$ at $\tau =$
0.5 fm and 1.5 fm respectively showing expansion of bubbles. (c),(d) 
show the plots of the phase of $l$ at $\tau = 0.5 $ and 3.2 fm. 
(e) and (f) show the surface plots of energy density (in GeV/fm$^3$) 
at $\tau = 0.75$ and 2.6 fm showing very different energetics of the walls 
of the true vacuum bubbles and the metastable vacuum bubbles.}
\label{Fig.3}
\end{figure*}

\subsection{variance of energy density}

 General evolution of bubble coalescence and formation of walls and strings
are similar to those shown in 
ref.\cite{gupta} for the $b_1 = 0$ case and we do not show those here.
As we are discussing the case of relatively small value of $b_1$ here 
we do not expect dramatic effects arising from explicit symmetry breaking 
(e.g. from the different mechanism of production of topological objects as 
demonstrated in ref.\cite{digal}). However, it is still important to see 
if there are any qualitative differences between the $b_1 = 0$ case and $b_1
\ne 0$ case. We find an interesting difference in the plot of the variance
of energy density between the two cases. We calculated the variance of 
energy density $\Delta \varepsilon$ at each time stage to study how energy 
fluctuations change during the evolution. In Fig.4 we show the plot of 
$\Delta \varepsilon/\varepsilon$ as a function of proper time. Here 
$\varepsilon$ is the average value of energy density at that time stage. 
The energy density $\varepsilon$ decreases due to longitudinal expansion, 
hence we plot this ratio to get an idea of relative importance of energy 
density fluctuations. For comparison we reproduce such a plot from
ref.\cite{gupta} for $b_1 = 0$ case in Fig.4b. We note that fluctuations 
have an overall tendency to decrease in Fig.4b while there seems no such
decrease in Fig.4a for the case with quark effects. Note also the presence
 of a peak for small times near $\tau \simeq$ 3 fm in $b_1 > 0 $ case. 
There is no such sharp peak for 
$b_1  = 0$ case. Remaining features of the plot can be interpreted as follows.
The initial rapid drop in $\Delta \varepsilon/\varepsilon$ is due to 
large increase in $\varepsilon$ during the heating stage upto $\tau$ =  
1 fm, followed by a rise due to increased energy density fluctuations 
during the stage when bubbles coalesce and bubble walls decay, as expected. 
The  peak in the plot near $\tau =$ 10 fm when $T$ drops below $T_c$
should correspond to the decay of domain walls and may provide a signal 
for the formation and subsequent decay of such objects in RHICE.

\begin{figure*}[!hpt]
\begin{center}
\leavevmode
\epsfysize=6truecm \vbox{\epsfbox{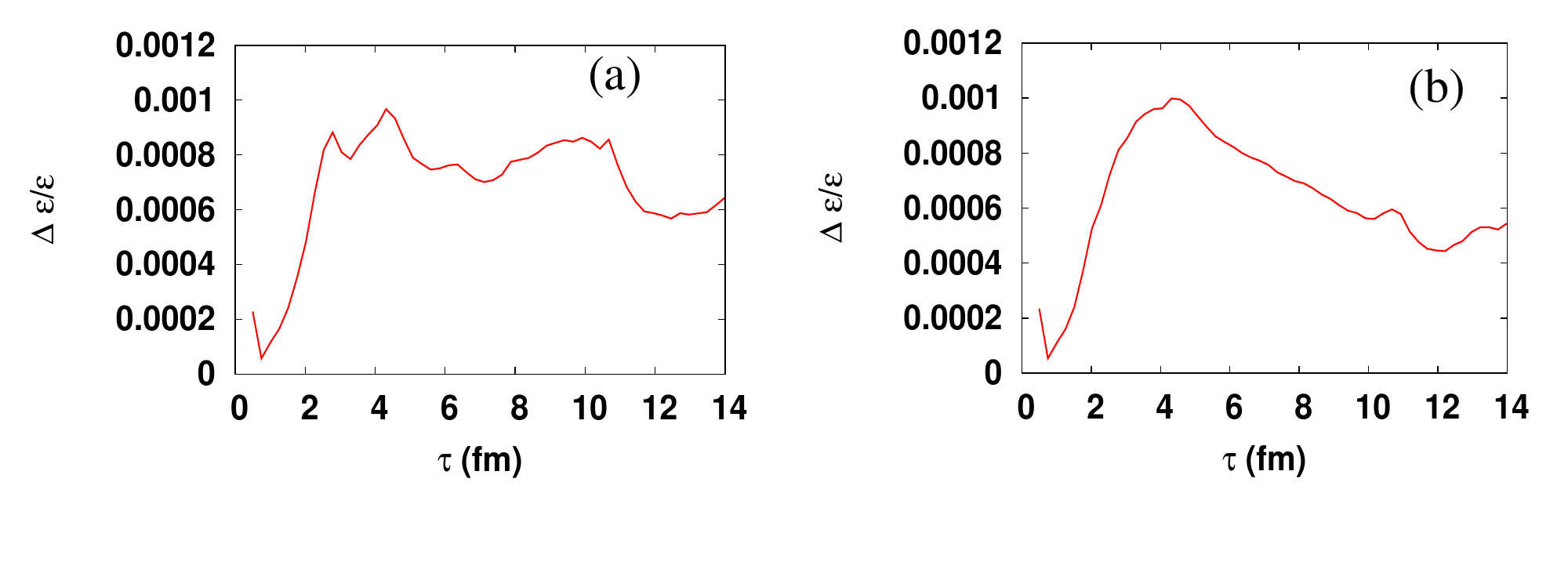}}
\end{center}
\caption{}{(a) and (b) show plots of the ratio of variance of energy 
density $\Delta \varepsilon$ and the average energy density $\varepsilon$ 
as a function of proper time for $b_1 = 0.005$ case and $b_1 = 0$ case
respectively.}
\label{Fig.4}
\end{figure*}

 The small peak at short times for $b_1 > 0 $ case seems to arise from
the difference between the collisions of metastable vacuum bubbles
and true vacuum bubbles and hence seems of qualitative importance. 
We have checked for various situations, different number of bubbles
etc. and this peak is always present. Fig.5 shows different cases,
for number of bubbles ranging from 4 to 10 and we see the presence of
this peak in all these cases. 

\begin{figure*}[!hpt]
\begin{center}
\leavevmode
\epsfysize=7truecm \vbox{\epsfbox{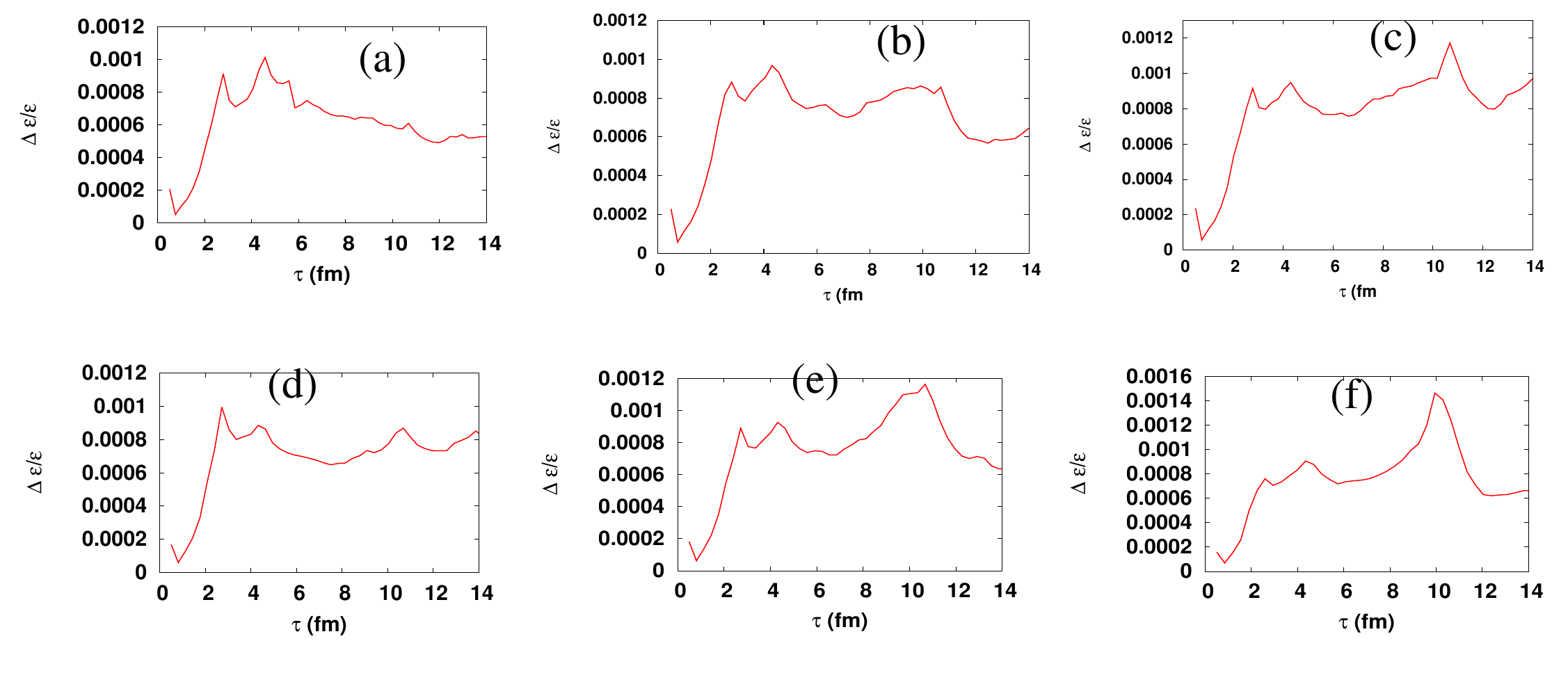}}
\end{center}
\caption{}{Plots of the ratio $\Delta \varepsilon / \varepsilon$ 
as a function of proper time for $b_1 = 0.005$ case for different
number of bubbles. (a) - (f) show curves for number of bubbles =
4,5,5(different realization),7,8, and 10, respectively. 
Note the presence of small peak for short times in all these cases.
Also note that there is no overall decrease for large times as was seen
in Fig.4b for $b_1 = 0$ case.}
\label{Fig.5}
\end{figure*}

\subsection{wall velocity}

 An important difference we note is in the wall velocity.
We have estimated wall velocities for the domain walls separating
the two degenerate metastable Z(3) vacua, and the metastable and the 
true vacuum. We find that the typical velocity of the domain walls
separating the two (degenerate) metastable vacua is 0.7 - 0.8,
similar to that obtained in ref. \cite{gupta} for $b_1 = 0$ case.
This is certainly expected. However, the velocity of domain wall
separating the true vacuum and the metastable vacuum is found
to be much larger in many cases, close to 1. Very accurate wall velocity
estimates are not possible due to uncertainties in identifying
wall location (with dynamically evolving wall profile). We show in 
Fig.6 and Fig.7 two different cases of 5 bubble nucleations (with different
locations and phases inside the bubbles). Contour plots of energy density 
are shown in Fig.6a and 6b at $\tau = 7.2$ and 7.8 fm (temperature
at these stages is 208 and 201 MeV respectively). The portion of
the domain wall near x = 14 fm, y = 12 fm in Fig.6a is seen to
move towards left in Fig.6b with $v \simeq 1$. This is confirmed
by the profile plot of $l_0 - l$ in Figs.6c and 6d at same 
stages, $\tau = 7.2$ and 7.8 fm respectively.

\begin{figure*}[!hpt]
\begin{center}
\leavevmode
\epsfysize=10truecm \vbox{\epsfbox{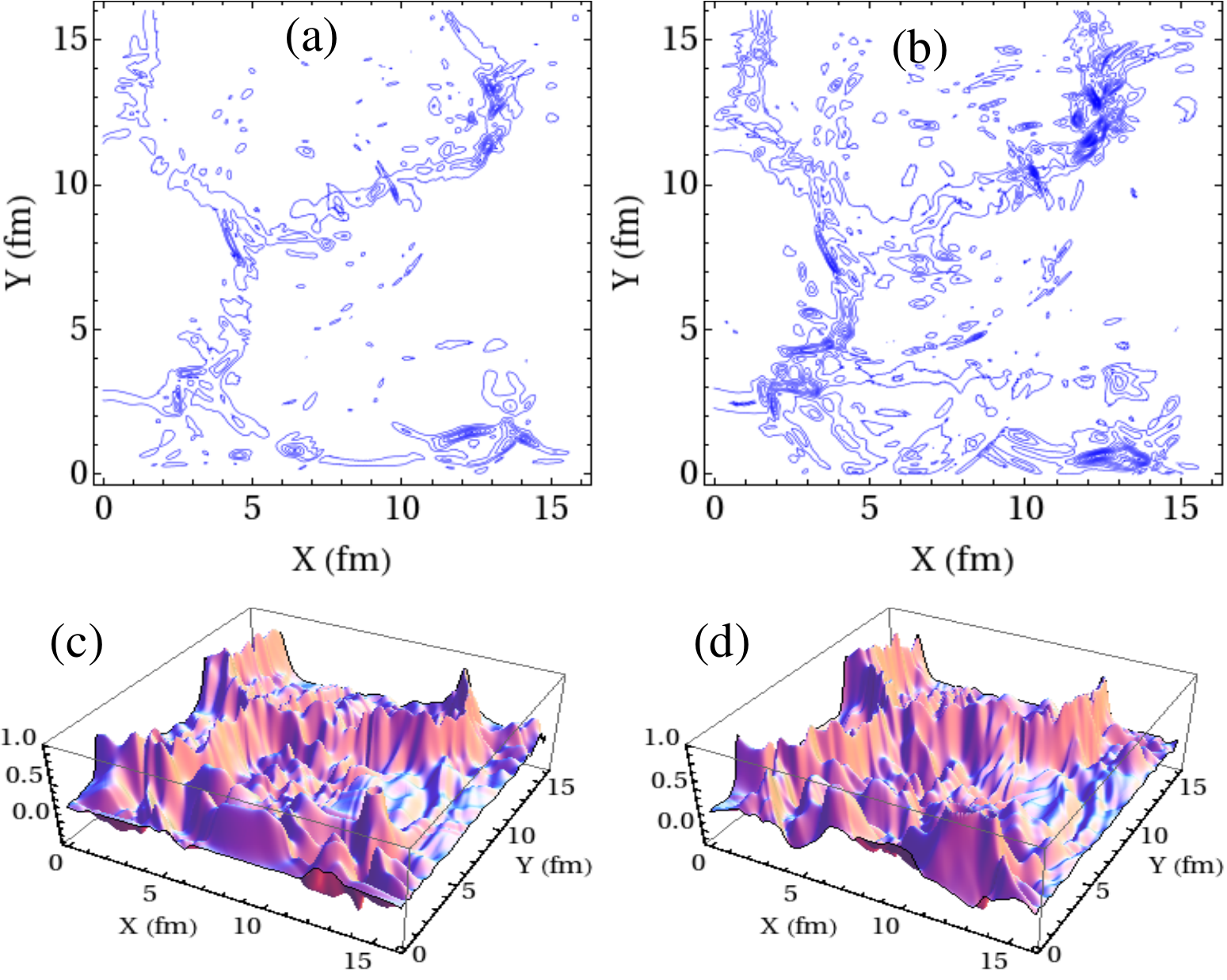}}
\end{center}
\caption{}{Contour plot of energy density at (a) $\tau = 7.2$ fm and
(b) at $\tau = 7.8$ fm. Wall portion near x=14 fm, y=12 fm
in (a) is seen to move towards left in (b) with large velocity.
(c) and (d) show the profile plots of $l_0 - l$ at these stages
confirming the motion of the domain wall.}
\label{Fig.6}
\end{figure*}

Fig.7 shows a different case of 5 bubble nucleation. (a) and (b)
show the contour plots of energy density at $\tau = 9.6$ and 10.9 fm.
Temperature at these stages is $T = $ 188 and 180 MeV (note, this is 
slightly below $T_c$). The location of wall in (a) is near x=9 fm,y=8 fm 
and this is seen to move towards lower right corner. This wall motion is 
confirmed by the profile plots of $l_0 -l$ in Figs.7c,d.

\begin{figure*}[!hpt]
\begin{center}
\leavevmode
\epsfysize=10truecm \vbox{\epsfbox{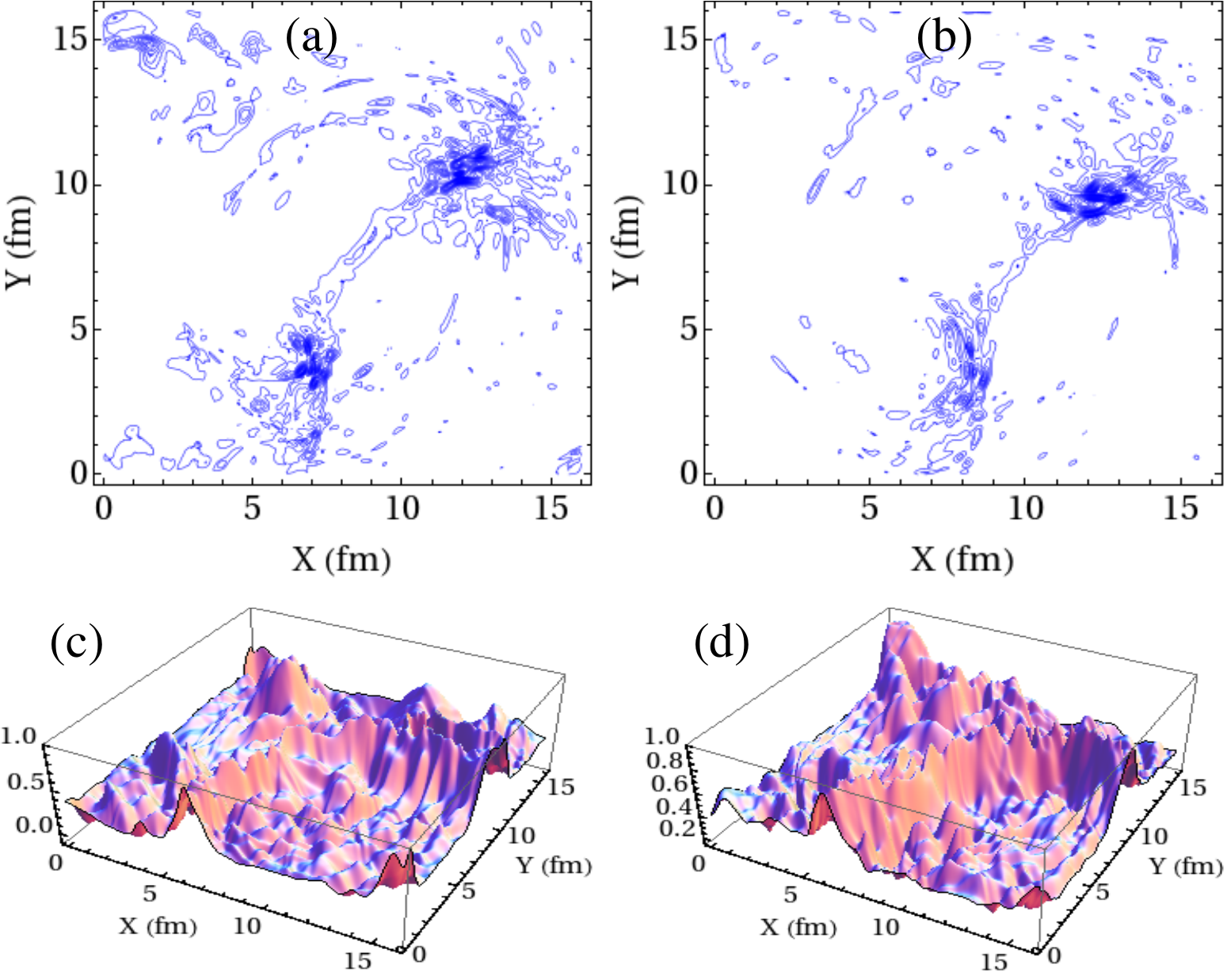}}
\end{center}
\caption{}{A different realization of 5 bubble nucleation.
Contour plot of energy density at (a) $\tau = 9.6$ fm and
(b) at $\tau = 10.9$ fm. Wall portion near x=9 fm, y=8 fm
in (a) is seen to move towards lower right in (b) with large velocity.
(c) and (d) show the profile plots of $l_0 - l$ at these stages
confirming the motion of the domain wall.}
\label{Fig.7}
\end{figure*}

\subsection{Rapid Collapse and Re-expansion}

Perhaps the most dramatic difference between the present case with
$b_1 \ne 0$ and the previous case \cite{gupta} of $b_1 = 0$ is seen
in Fig.8 and Fig.9. This shows the case of nucleation of 10 bubbles
in a region of 22 fm $\times$ 22 fm. Though both of these numbers
are somewhat large for RHICE, at least the size may not be too 
unrealistic for later stages of plasma evolution.  Fig.8 shows a time 
sequence of the contour plot of energy density at $\tau = $ 6.44, 8.20, 9.94,
11.34, 12.74, and 14.5 fm. The temperature at these stage is $T =$
215.0, 198.4, 186.0, 178.0, 171.2, and 164.1 MeV respectively.
Note that $T$ is below in (d) and after that. A closed domain wall
is seen in the lower left region in (a) with x = 1-6 fm and 
y = 5-11 fm. This wall collapses rapidly by $\tau = 9.94$ fm.
The collapse velocity again is seen to be close to $v \simeq 1$.
Interesting dynamics is seen for later plots when an expanding
front is seen from the point of collapse. It rapidly expands,
again with $v \simeq 1$ all the way until last stages in (f).
Presence of such an energetic expanding front is confirmed
by the surface plots of energy density at the same stages as
shown in Fig.9. Due to very large velocity and sharp profile of
the expanding front it may well represent a shock front in
the plasma.

\begin{figure*}[!hpt]
\begin{center}
\leavevmode
\epsfysize=10truecm \vbox{\epsfbox{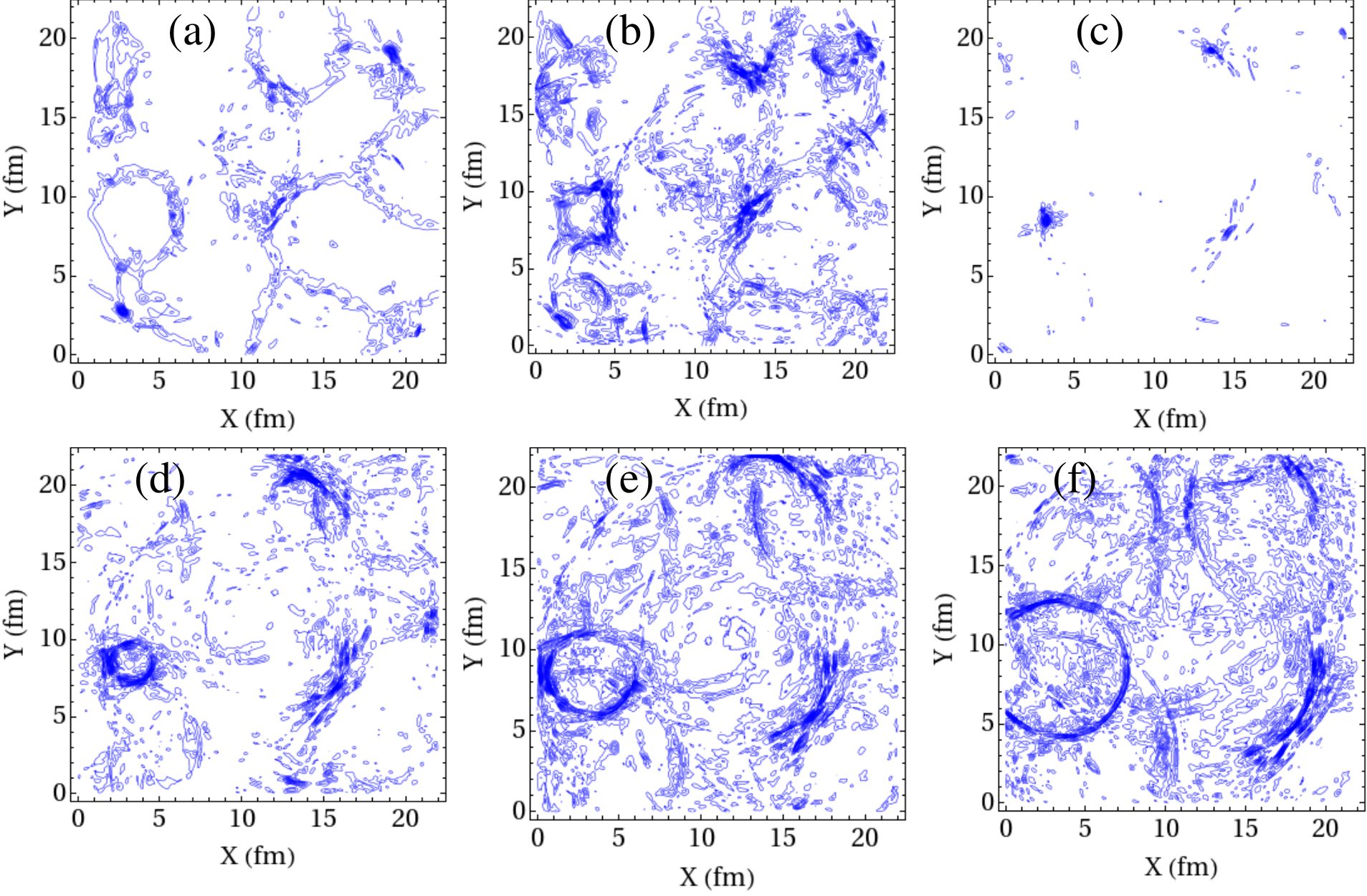}}
\end{center}
\caption{}{A case of 10 bubbles in 22 fm $\times$ 22 fm region.
This figure shows a time sequence of contour plot of energy density showing
rapid collapse of a domain wall (towards lower left) and subsequent
rapid expansion of a circular front.}
\label{Fig.8}
\end{figure*}

\begin{figure*}[!hpt]
\begin{center}
\leavevmode
\epsfysize=8truecm \vbox{\epsfbox{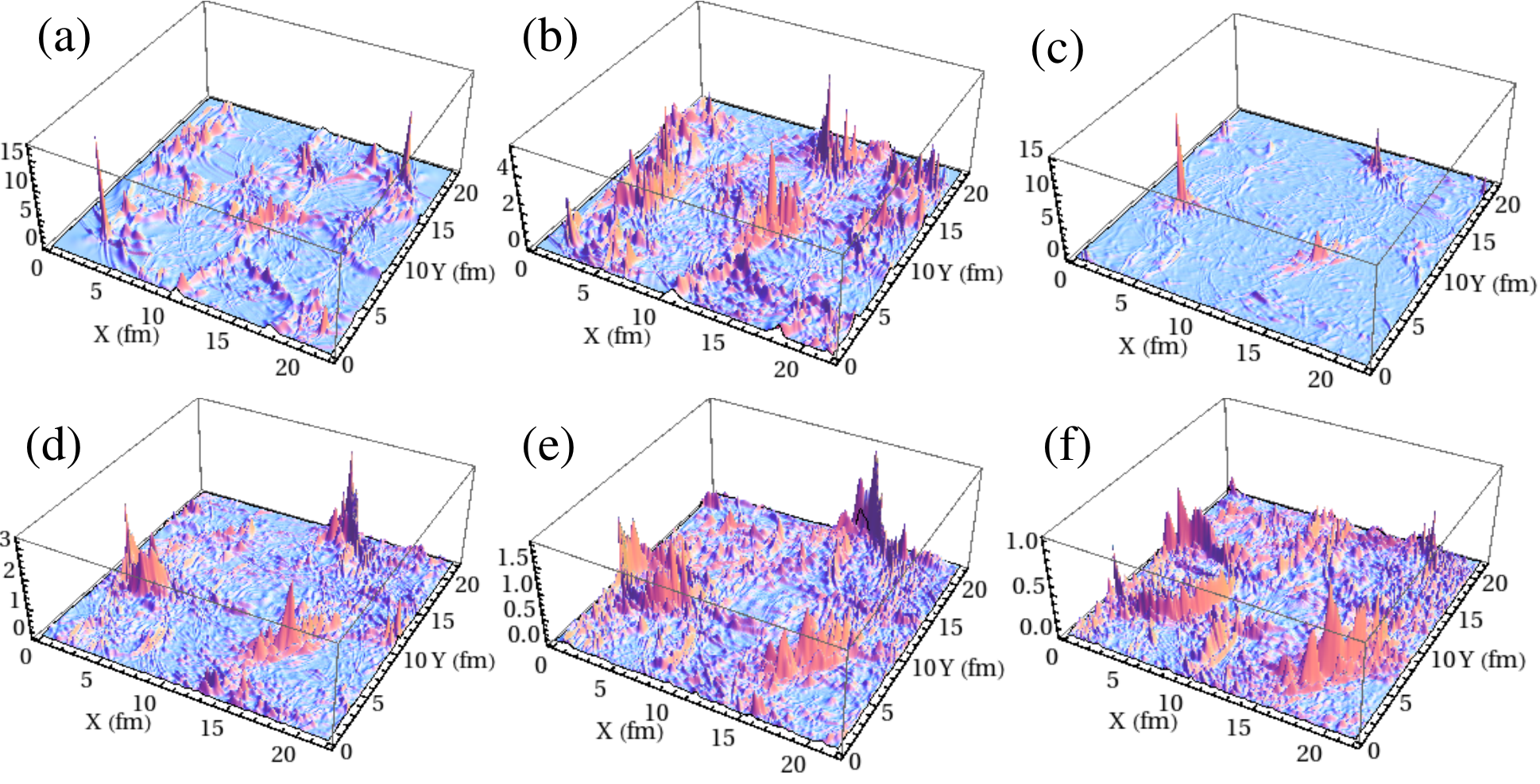}}
\end{center}
\caption{}{Surface plots of energy density for various stages shown
in Fig.8}
\label{Fig.9}
\end{figure*}

\section{Possible experimental signatures of Z(3) walls and strings
with explicit symmetry breaking}
\label{sec:signtr}

 The $Z(3)$ wall network and associated strings exist only during
the QGP phase, melting away when the temperature drops below $T_c$.
However, they may leave their signatures in the distribution of 
final particles due to large concentration of energy density in 
extended regions as well as due to non-trivial scatterings of quarks 
and antiquarks with these objects.  The extended regions of high energy 
density resulting from the domain walls and strings are clearly seen in
our simulations and some extended structures/hot spots also survive
after the temperature drops below the transition temperature $T_c$.
This is just as was seen in the case of $b_1 = 0$ case in ref.\cite{gupta}. 
We again mention that even the hot spot resulting from the collapse of closed
domain walls in our simulations will be stretched in the longitudinal 
direction into an extended linear structure (resulting from the collapse of
a cylindrical wall). These may be observable in the analysis of
particle multiplicities. This is important especially in respect
to the ridge phenomenon seen at RHIC \cite{ridge}. In view of lasting
extended energy density fluctuations from Z(3) walls, it is of interest
to check if these structures can account for the ridge phenomenon.

  Our results show interesting pattern of the evolution of  the 
fluctuations in the energy density which show that these fluctuations
do not decrease with time which was the case for $b_1 = 0$ case
studied in ref. \cite{gupta}. Especially important may be the presence of 
small additional peak of short times for $b_1 > 0$ case. Fluctuations near 
the transition stage may leave direct imprints on particle 
distributions. However, dileptons or direct photons should be sensitive
to these fluctuations, and these may give a time history of
evolution of such energy density fluctuations during the early
stages. In such a case the existence of small peak for $b_1 > 0$ case
may be observable. 

 A dramatic difference between the case of $b_1 = 0$ and $b_1 \ne 0$
is seen in Figs.8,9. Collapse of a closed wall is expected and was seen
for $b_1 = 0 $ case also, though the wall speed here is much higher,
close to 1. In general we have seen here that walls separating 
true vacuum from metastable vacuum have speeds much higher than seen
for the case of $b_1 = 0$. What is qualitatively new in the present
case is rapidly expanding circular front after the collapse of the wall.
This front continues its speed and shape even when temperature
drops below $T_c$. Possibility of such expanding circular (cylindrical,
with longitudinal expansion) energetic
fronts should have important implications on particle momenta,
especially on various flow coefficients.

  Another important difference due to $b_1 > 0$ is expected in
investigating the  interactions of quarks and antiquarks 
with domain walls. Earlier we had argued \cite{gupta} that collapsing
Z(3) walls will lead to concentration of quarks (due to small
non-zero chemical potential in RHICE) in small regions \cite{znb}. This will
lead to enhancement of baryons, especially at high $P_T$ \cite{apm} due
to $P_T$ enhancement of quarks/antiquarks as they undergo repeated 
reflections from a collapsing wall. 
(There is also a possibility of spontaneous CP violation in the scattering
of quarks and antiquarks from Z(3) walls, see ref.\cite{cpvln}.) However, 
with $b_1 >0$, there may also be a possibility that some Z(3) wall may actually 
expand (the one enclosing the true vacuum and with sufficiently large size).
In that case it will have opposite effect and baryon number will
be more diffused.  Even the enhancement of $P_T$ may happen for some
domain walls (those which enclose metastable vacuum) while the expanding
closed walls (enclosing the true vacuum) should lead to the redshift
of the momenta for the enclosed quarks. 
All these issue need to be explored with more elaborate
simulations. In this context the difference in the wall velocity between 
different types of Z(3) walls is of importance.  While studying the effects 
of quark reflections from these walls and associated modification of 
$P_T$ spectrum, wall velocity is of crucial importance and the presence
of different types of collapsing Z(3) walls may lead to bunches of
hadrons with different patterns of modified $P_T$ spectra.

\section{Conclusions}

  We have studied the effects of explicit symmetry breaking arising
from quark effects on the  formation and evolution of $Z(3)$ 
interfaces and associated strings. Explicit symmetry breaking makes
Z(3) vacua non-degenerate with two vacua $l = z, z^2$ remaining degenerate
with each other but having higher energy than the true $l = 1$
vacuum. Thus $l = z, z^2$ vacua become metastable. We have used an 
effective potential for the Polyakov loop expectation value $l(x)$ from ref. 
\cite{psrsk,psrsk2} with incorporation of explicit symmetry breaking
in terms of a linear term in $l$ and have studied the dynamics
of the (C-D) phase transition in the temperature/time range when the
first order transition of this model proceeds via bubble nucleation.
This allows for only relatively small explicit symmetry breaking 
(characterized  by the strength $b_1$ of the linear term in $l$).
We again emphasize that, though our study is in 
the context of a first order transition, its results are expected to 
be valid even when the transition is a cross-over. This is because
our focus is only on the formation of topological objects whose
formation (via Kibble mechanism) only depends on the formation of
a domain structure and not crucially on the dynamics of the phase
transition. Though, our statements about the energetics of bubble walls
etc. clearly apply only for a first order transition.

  An important result we have discussed in this paper relates to 
expected relative importance of the metastable Z(3) vacua. Due
to higher energy of these vacua one would expect that bubbles
with these vacua should form with relatively lower probability
(even with small values of $b_1$ we have used). However, we find
interesting results due to nontrivial interplay of the pre-exponential
factor and the exponential term in the nucleation rate for the
bubbles. While the exponential term leads to a decrease in the rate
for metastable vacua due to larger action, the pre-exponential factor
leads to an increase in the rate for larger action. For a suitable
range of temperatures, which for our choice of parameter values
lies between $T = 200 $ MeV to $T = 225$ MeV, the metastable vacuum
bubbles have the same or larger nucleation rate compared to the true
vacuum bubbles. As the metastable vacuum bubbles also have larger
sizes it means that a larger fraction of QGP phase may get converted
to the metastable Z(3) vacua than to the true Z(3) vacuum. The dynamics
of these domains being so different its effects on the evolution of
plasma and various signals may be important.

\acknowledgments

  We are very grateful to Sanatan Digal, Anjishnu Sarkar, Ananta P. Mishra,
P.S. Saumia, and Abhishek Atreya for very useful comments and suggestions.
USG, AMS, and VKT acknowledge the support of the Department of Atomic 
Energy- Board of Research in Nuclear Sciences (DAE-BRNS), India, under 
the research grant no 2008/37/13/BRNS. USG and VKT acknowledge support of 
the computing facility developed by the Nuclear-Particle Physics group
of Physics Department, Allahabad University under the Center of
Advanced Studies (CAS) funding of UGC, India.


\end{document}